\documentclass[12pt]{article}
\usepackage{epsfig}

\RequirePackage{color}

\title{\bf
Scaling behaviour of lattice animals at the upper critical dimension}
\author{ 
{\it Christian~von Ferber$^{\,1}$,} 
{\it Damien Foster$^{\,2}$,}
{\it Hsiao-Ping~Hsu$^{\,3}$}  
and 
{\it Ralph~Kenna$^{\,1}$}\\~\\
$^1$ Applied Mathematics Research Centre,
Coventry University,\\
Coventry, CV1 5FB, England
{}\\~\\
$^2$ 
Laboratoire de Physique Th{\'{e}}orique et Mod{\'{e}}lisation (CNRS UMR 8089), \\
Universit{\'{e}} de Cergy-Pontoise, \\
2 ave A. Chauvin 95302 Cergy-Pontoise cedex, France
{}\\~\\
$^3$ Institut f\"ur Physik, Johannes Gutenberg-Universit\"at Mainz\\
D-55099 Mainz, Staudinger Weg 7, Germany
 }
\begin{document}

\maketitle
                      {\Large
                      \begin{abstract}
We perform numerical simulations of the lattice-animal problem
at the upper critical dimension $d=8$ on  hypercubic
lattices in order to investigate logarithmic corrections to scaling there.
Our stochastic sampling method is based on the pruned-enriched 
Rosenbluth method (PERM), appropriate to linear
polymers, and yields high statistics with animals comprised of up to 8000 sites.
We estimate both the partition sums (number of different animals) and the radii of gyration. 
We re-verify the Parisi-Sourlas prediction for the leading exponents
and compare the logarithmic-correction exponents to two
partially differing sets of predictions from the literature.
Finally, we propose, and test,  a new Parisi-Sourlas-type scaling relation appropriate for the logarithmic-correction exponents.
%
                        \end{abstract} }
%
  \thispagestyle{empty}
%
%
  \newpage
%
                  \pagenumbering{arabic}

\section{Introduction}

A lattice animal is a cluster of connected sites on a regular lattice. 
The enumeration of such objects -- also called  polyominoes -- is a combinatorial problem of interest
to mathematicians \cite{Golomb94}, while in physics, they are closely linked to the problems of percolation \cite{Stauffer92} and clustering in spin models \cite{FoKa72}.
In chemistry they form a basis for models of randomly branched polymers in  good solvents \cite{LuIs79}.
Lattice animals linked by translations are considered as belonging to the same equivalence class, and as such are considered to be essentially the same.
Of interest is  $Z_N$, the number of distinct animals containing $N$ sites. A related objective is the 
calculation of the radius of gyration $R_N$, related to the average distance of  occupied sites from the centre of mass of the lattice animal. A number of variants of the lattice animal are studied: bond lattice animals, which are clusters of connected bonds; weakly embedded and strongly embedded trees. It is believed that all these different models belong to the same universality class \cite{HsNa05}. 

It is now established that the number of lattice animals and the radius of gyration behave, to leading order in $N$, as
\cite{LuIs79} 
\begin{eqnarray}
Z_N & \sim & \mu^{N}N^{-\theta} \,, 
\label{1}\\
R_N & \sim & N^{\nu}\,.
\label{2}
\end{eqnarray}
Here $\theta$ is related to the rate of growth of $Z_N$
and $\nu=1/d_H$ where $d_H$ is the Hausdorff (fractal) dimension of the lattice animals.

It is  useful to define the generating function (or grand-canonical partition function) for the model as
\begin{equation}\label{gen_func}
Z=\sum_{N=0}^\infty K^N Z_N\,.
\end{equation}
The growth constant $\mu$ is  related to the critical fugacity $K_c$ defining the radius of convergence of the generating function (\ref{gen_func}). To leading order,
\begin{equation}
Z\sim |1-\mu K_c|^{\theta-1},
\end{equation}
giving the identification $\mu=1/K_c$.
The growth constant $\mu$ is related to the critical fugacity 
of the corresponding field theory and depends on the lattice coordination number, while the entropic exponent $\theta$ and the correlation-length exponent $\nu$ are universal \cite{LuIs78}.

Lattice animals may be viewed as the graphs arising from high-temperature expansions of related magnetic models, in particular the high temperature expansion of (the derivative of) the free energy of the $q=1$ Potts models, which in turn is related to the percolation problem.  Parisi and Sourlas related the problem of branched polymers, and hence lattice animals, in $d$ dimension with the Yang-Lee edge problem in $d-2$ dimension, and predicted that $\theta$ and $\nu$ are  related by 
\cite{PaSo81}
\begin{equation}
 \theta = (d-2)\nu +1 \,.
 \label{PS}
\end{equation}
This relation was re-derived in an interesting fashion \cite{FFamily82}: identifying $Z$ as the high-temperature expansion of a magnetic model leads to the identification of $K$ with the magnetic and thermal scaling fields of the related magnetic model, indicating that the lattice animal is controlled by a single scaling field. This leads to the relations $3-\theta=\gamma=\alpha$. The exponents $\alpha$ and $\gamma$ are the usual critical indices  related to the divergence of the specific heat and susceptibility of the related Potts model. Substituting the mean-field exponents into the usual  hyperscaling relation $\alpha=2-d\nu$ would lead to an upper critical dimension of $6$, whereas the correct upper critical dimension is $8$, indicating that there is an anomalous scaling and that  hyperscaling is modified, with $d$ replaced
by $d-2$:
\begin{equation}
\alpha=2-(d-2)\nu
\end{equation}
and Eq.(\ref{PS}) is recovered.

In Ref.~\cite{HsNa05}, the Parisi-Sourlas predictions for the leading behaviour for both Eqs.(\ref{1}) and (\ref{2}) were verified in dimensions $d=2$ to $d=9$ using a 
high-statistics numerical study with lattice animals with up to several thousand sites in each case. 
The measured values of $\theta$ and $\nu$ were compatible with the Parisi-Sourlas scaling relation. 

Although experimentally inaccessible, a complete understanding of the lattice-animal problem includes 
 the upper critical dimension $d=8$.
At and above this dimension, the critical
exponents take on the mean-field values $\nu =1/4$ and $\theta = 5/2$ \cite{AdMe88}. 
In eight dimensions, the scaling forms (\ref{1}) and (\ref{2}) are modified by multiplicative logarithmic corrections. 
Indeed, in the high-precision study of Ref.~\cite{HsNa05}, very large corrections
to Eqs.(\ref{1}) and (\ref{2}) were reported in eight dimensions.
While it was presumed that these corrections are logarithmic in nature,
no attempt at a detailed fit to them was made because the authors were unaware of
theoretical predictions beyond the leading order, and because of the notorious difficulty
in fitting to such logarithms.
It is expected that at the upper critical dimension $d_c=8$, $Z_N$ and $R_N$  scale as
\begin{eqnarray}
Z_N & \sim & \mu^{N}N^{-\theta} (\ln{N})^{\hat{\theta}} \,,
\label{3} \\
R_N & \sim & N^{\nu} (\ln{N})^{\hat{\nu}} \,,
\label{4}
\end{eqnarray}
with $\theta=5/2$ and $\nu=1/4$. The values of the logarithmic correction exponents $\hat{\theta}$ and $\hat{\nu}$ 
are the subject of the present article.

The mean-field exponents for the lattice animal model correspond to the exponents calculated from a $\phi^3$ theory with reduced temperature $t=0$ and where the reduced magnetic field $h$ is used as a temperature-like variable. This is consistent with the realisation, stated above, that in this model there is only one scaling field, linked to the magnetic field of the underlying magnetic model. The full set of mean-field exponents are
\begin{equation}
\alpha=\frac{1}{2},\ \ \ \beta=\frac{1}{2},\ \ \ \gamma=\frac{1}{2},\ \ \ \delta=2,\ \ \ \nu=\frac{1}{4},\ {\rm and}\ \ \eta=0\,.
\end{equation}
These exponents are  related by Fisher renormalisation to the standard mean-field exponents obtained setting $h=0$ and varying $t$. These Fisher-renormalised exponents are
\begin{equation}
\alpha_X=-1,\ \ \ \beta_X=1,\ \ \ \gamma_X=1,\ \ \ \delta_X=2,\ \ \ \nu_X=\frac{1}{2},\ {\rm and}\ \ \eta_X=0\,.
\end{equation}
Whilst in the lattice animal model there is only a single scaling field ($h$), in the equivalent Yang-Lee model 
it is possible to vary both $t$ and $h$ independently. 
  
This pairing via Fisher renormalisation and scaling relations have permitted 
new analytic predictions for the logarithmic corrections \cite{us}: $\hat{\theta}=1/3$, consistent with the prediction of Ref.~\cite{RL} and $\hat{\nu}=1/9$. 
The latter differs from a previous renormalisation-group based prediction: $\hat{\nu}=-1/72$ \cite{RL}.
We therefore considered it worthwhile to revisit the problem of lattice animals in $d=8$ dimensions in an attempt to discern whether the numerics support either of these analytic predictions for logarithmic corrections.

In what follows we find numerical support for logarithmically-corrected scaling behaviour in eight dimensions,
with $\hat{\theta}=1/3$.
Although the numerics for the radius of gyration yield less convincing results, they appear more compatible with the value
$\hat{\nu}=1/9$ predicted in Ref.~\cite{us} than $\hat{\nu}=-1/72$ predicted in Ref.~\cite{RL}.

\section{Scaling at the upper critical dimension}


While the leading exponents $\theta$ and $\nu$ in Eqs.(\ref{3}) and (\ref{4}) are not in doubt,
there are two sets of predictions in the literature for their logarithmic-correction 
counterparts $\hat{\theta}$ and $\hat{\nu}$. 
In Ref.~\cite{RL}, Ruiz-Lorenzo analytically studied these and other logarithmic corrections
for a generic $\phi^3$ scalar field  theory at its upper critical dimension $d=6$. 
This theory, with imaginary coupling, is known to describe the Yang-Lee problem \cite{Fi78}.
The latter originates from the study of the Yang-Lee edge singularity, which may be regarded
as a critical or pseudo-critical point. 

Parisi and Sourlas advanced a relationship between the Yang-Lee singularity in $D$ dimensions
and the lattice-animal problem in $d=D+2$ dimensions \cite{PaSo81}.
Recently, an exact mapping between the two problems established this relationship 
on a rigorous footing~\cite{rig}.
The renormalisation-group calculation of Ref.~\cite{RL} for the Yang-Lee problem 
($\phi^3$ theory with imaginary coupling) in $d=6$ dimensions yields a free energy 
as a function of the magnetic field, the singular part of which is of the form
\begin{equation}
 f \sim h^{\frac{3}{2}} (\ln{h})^\frac{1}{3}\,.
 \label{RLf}
\end{equation}
The grand canonical partition function for the lattice animals is 
\begin{equation}
 Z = \sum_NK^NZ_N\,,
 \label{Z}
\end{equation}
where, with  $Z_N$  given by Eq.(\ref{4}), scales as 
\begin{equation}
 Z \sim |1-K\mu|^{\theta-1}|\ln{|1-K \mu|}|^{\hat{\theta}}\,.
 \label{ZZ}
\end{equation}
The Parisi-Sourlas mapping, then, identifies Eq.(\ref{RLf}) for the Yang-Lee problem with 
Eq.(\ref{ZZ}) for the lattice animals, with the magnetic field in the former case
being replaced by the fugacity in the latter.
This leads to the predictions $\theta = 5/2$ and $\hat{\theta}=1/3$
for the lattice-animal problem in $d=8$ dimensions \cite{RL},
which are supported by previous direct calculations  \cite{LuIs79}.
The $\phi^3$-approach also leads to analytic predictions for the correlation length,
\begin{equation}
\xi(h) \sim h^{-1/4} |\ln{h}|^{-1/72},
\end{equation}
 which translates to 
$\nu = 1/4$ and $\hat{\nu}=-1/72$ for lattice animals.

The lattice animal and Yang-Lee problems can be considered either 
with the field or the order parameter held constant, with 
constant field being the more natural in field theory \cite{LuMc81}.
The corresponding two sets of critical exponents are linked
via Fisher renormalization \cite{Fi68}.
The Fisher renormalization scheme for logarithmic-corrections
was recently established in Ref.~\cite{us}.
In Ref.~\cite{RL}, Ruiz-Lorenzo has also determined the 
constant-order-parameter critical exponents and,
in particular, the constant-order-parameter logarithmic-correction
exponent for the  correlation-length  is
given as $\hat{\nu}_X=5/18$ \cite{RL}.
Applying the Fisher-renormalization relations from Ref.~\cite{us}
with this value yields the prediction $\hat{\nu} = 1/9$,
which is {\emph{different}} to the constant-field estimate
$\hat{\nu}=-1/72$ quoted above.

To summarize, the renormalization-group approach of Ref.~\cite{RL} yields
constant-field estimates for the leading entopic and Flory critical exponents 
which agree with mean-field, have been checked numerically in Ref.~\cite{HsNa05}
and which are not in doubt.
The corresponding logarithmic exponents are $\hat{\theta}=1/3$ and $\hat{\nu}=-1/72$.
In Ref.~\cite{RL}, constant-order-parameter estimates are also given, which,
when Fisher renormalized also yield $\hat{\theta}=1/3$ but $\hat{\nu}=1/9$.
while all leading exponents and Ruiz-Lorenzo's RG calculations for the $\phi^3$ theory agree with mean-field predictions,
and while  calculations for the logarithmic-correction exponents agree with all previous estimates
where they exist (besides the Yang-Lee problem, these include for 
spin glasses and for percolation in six dimensions),
the disparity between the estimates for $\hat{\nu}$ requires
further investigation and we chose a non-perturbative, numerical approach. 
 It is also necessary to check 
if  $\hat{\theta}=1/3$ is supported numerically, as this has not been 
tested non-perturbatively before.

\section{Numerical approach}

The numerical data was obtained using  the prune-enriched Rosenbluth method (PERM) which is a variant of the Rosenbluth-Rosenbluth Monte Carlo 
method for self-avoiding walks designed to avoid the ensembles being dominated by a few high weight clusters and to avoid undue time being used calculating clusters with small weights \cite{Care}.
This is avoided by reducing 
the width of the weights distribution by pruning low-weight configurations 
while cloning high-weight ones. To apply this approach to lattice animals, we have
to estimate the cluster weight while it is still growing. 
The approach, which is discussed in detail in Ref.~\cite{HsNa05}, generates independent clusters from different 
Monte Carlo tours and therefore leads to straightforward estimates for the errors
in the raw data for the partition sums and  gyration radii.

\begin{figure}[!t]
\begin{center}
\includegraphics[width=11cm,clip]{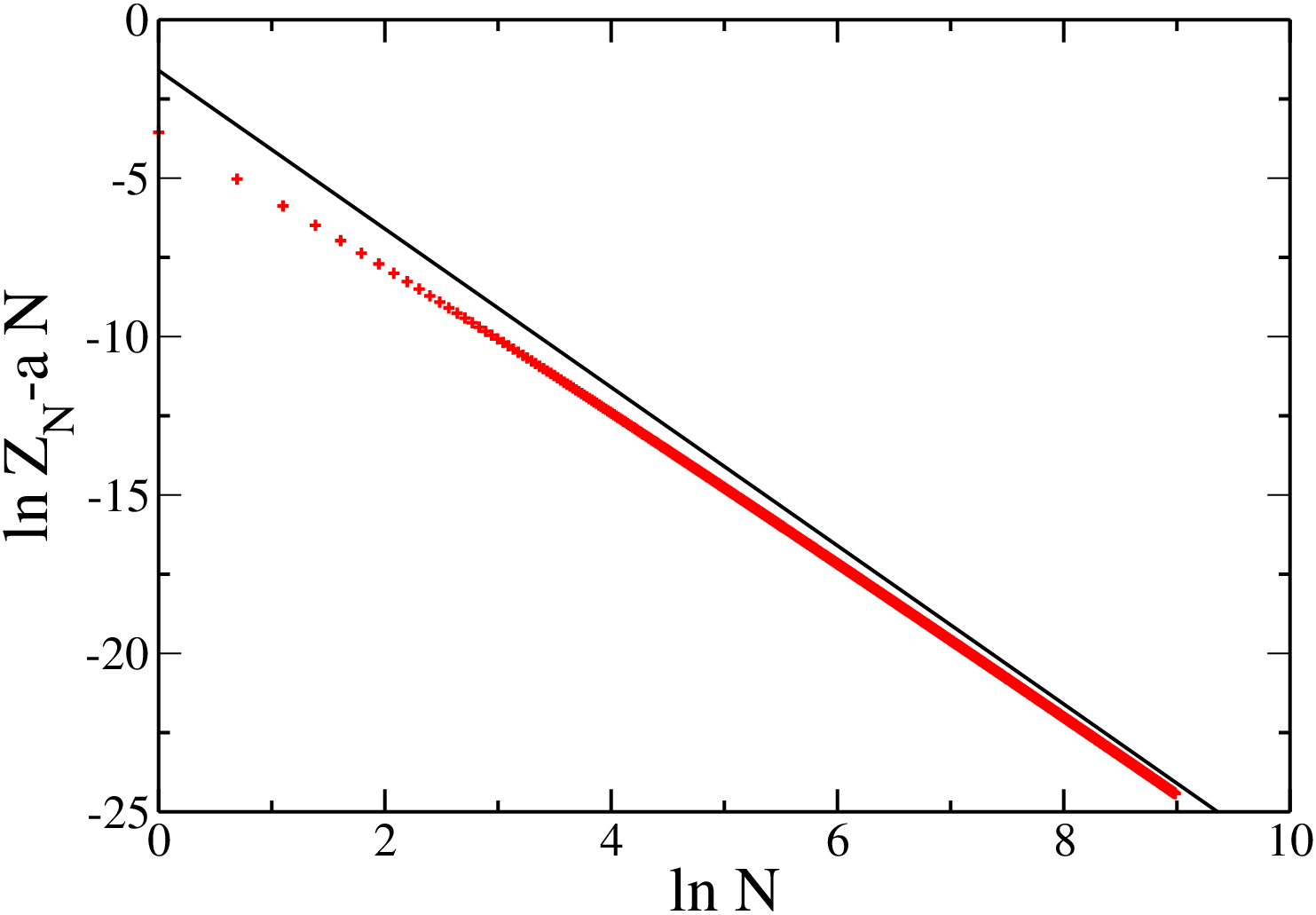}
\caption{The leading dependency of the partition sum $Z_N$ on the animal size $N$
with $a = \ln{\mu} =3.554830$. The line is of slope $-5/2$ to guide the eye.
}
\label{Zn}
\end{center}
\end{figure}

\begin{figure}[!t]
\begin{center}
\includegraphics[width=11cm,clip]{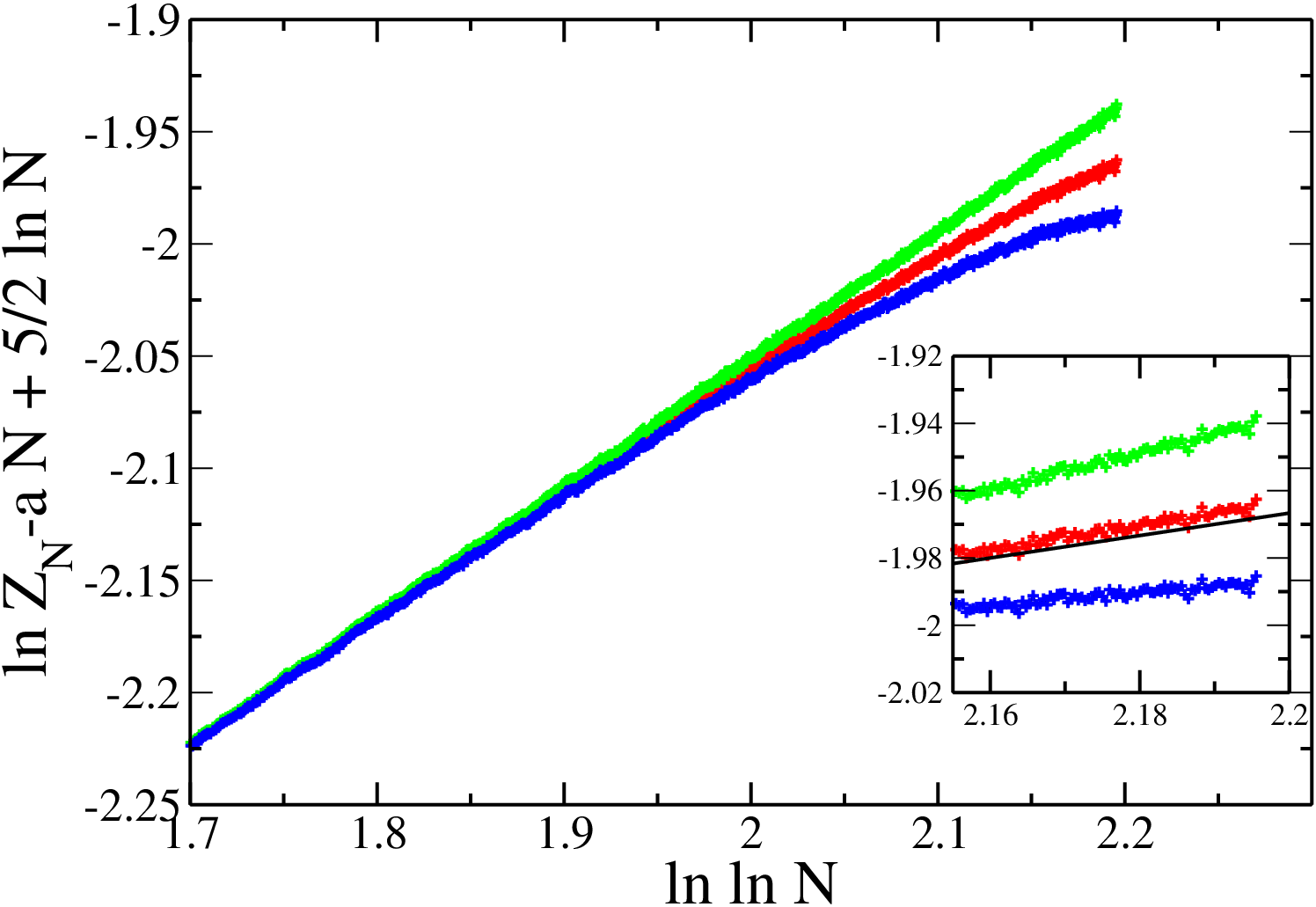}
\caption{
The three sets of data correspond to 
$a = \ln{\mu} =3.554827$ (upper set)
$a =3.554830$ (middle)
$a =3.554833$ (lower).
The asymptotic slope is expected to yield the 
logarithmic-correction exponent $\hat{\theta}$. 
While the pre-asymptotic, smaller-$N$ data 
has a slope of about $0.58$, the middle data of the inset, which corresponds to larger $N$ values  
has a line of slope $\sim 1/3$ and
a line of this slope is included  to guide the eye.
}
\label{Zncorr}
\end{center}
\end{figure}

In Fig.~\ref{Zn} $\ln Z_N-a N$ is plotted against $\ln N$. The constant $a=\ln\mu$ is set to the best fit value $a = \ln{\mu} =3.554830$. The asymptotic value of the slope is consistent with the expected value of $\theta=5/2$. 
To investigate the logarithmic corrections, the leading scaling behaviour is subtracted out, and  
$\ln{Z_N} - aN + (5/2) \ln{N} $ is plotted against $\ln{(\ln{N})}$
in Fig.~\ref{Zncorr}. The middle line shows the plot corresponding to the best fit value of $a=3.554830$, whilst the other two lines correspond to the upper and lower error bounds $a=3.554827$ (upper line) and $a=3.554833$ (lower line). The insert shows the last section of the curve, which can be seen to have a slope consistent with $\hat{\theta}=1/3$. 
(Note that although the horizontal axis has relatively short range, since it is 
on a log-log scale it corresponds to a wide range of animals sizes,
from  $N \approx 6000$ to $N = 8000$.)

\begin{figure}[!t]
\begin{center}
\includegraphics[width=11cm,clip]{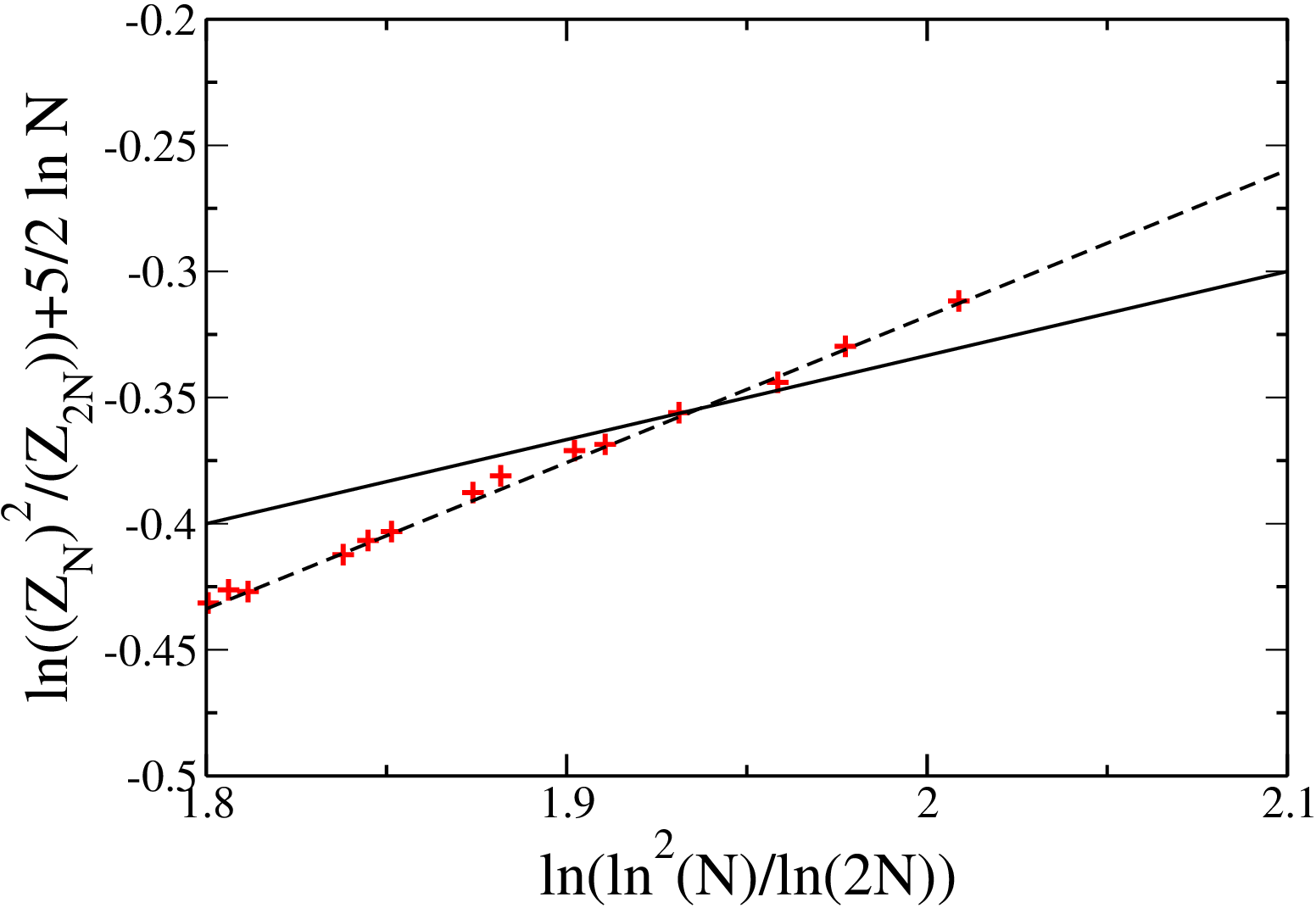}
\caption{The points show the log correction to $Z^2_N/Z_{2N}$. The solid line  has the expected slope of 
$\hat{\theta}=1/3$.
The dashed line is a fit to the data and has slope $0.58$. The measured slope corresponds to the slope of the pre-asymptotic part of the curve shown in Fig.~\ref{Zncorr} (see text).
}
\label{HP}
\end{center}
\end{figure}

The slope in the sizes of animals calculated is sensitive to the precise value of $a$. 
One may attempt to  eliminate $a$ by  using two values of $N$, as
\begin{equation}
\frac{Z_N^2}{Z_{2N}} \sim N^{-\theta}\left(\frac{\ln^2 N}{\ln 2N}\right)^{\hat{\theta}}\,.
\end{equation}
In Fig.~\ref{HP} we plot  $\ln(Z_N^2/Z_{2N}) + 5/2\ln N$ against $\ln (\ln^2 N/\ln 2N)$, and the larger-$N$ value of the slope appears closer to $0.58$ than to the expected $\hat{\theta}=1/3$. 
However, this method relies heavily on the first half of the data, which corresponds to 
relatively small $N$ values and is far from asymptotic. 
Indeed, the pre-asymptotic portion of the best-fit curve in Fig.~\ref{Zncorr} is also well fitted by a straight line of slope $0.58$, and we therefore consider the asymptotic regime
not to have been reached in Fig.~\ref{HP}.

\begin{figure}[!t]
\begin{center}
\includegraphics[width=11cm,clip]{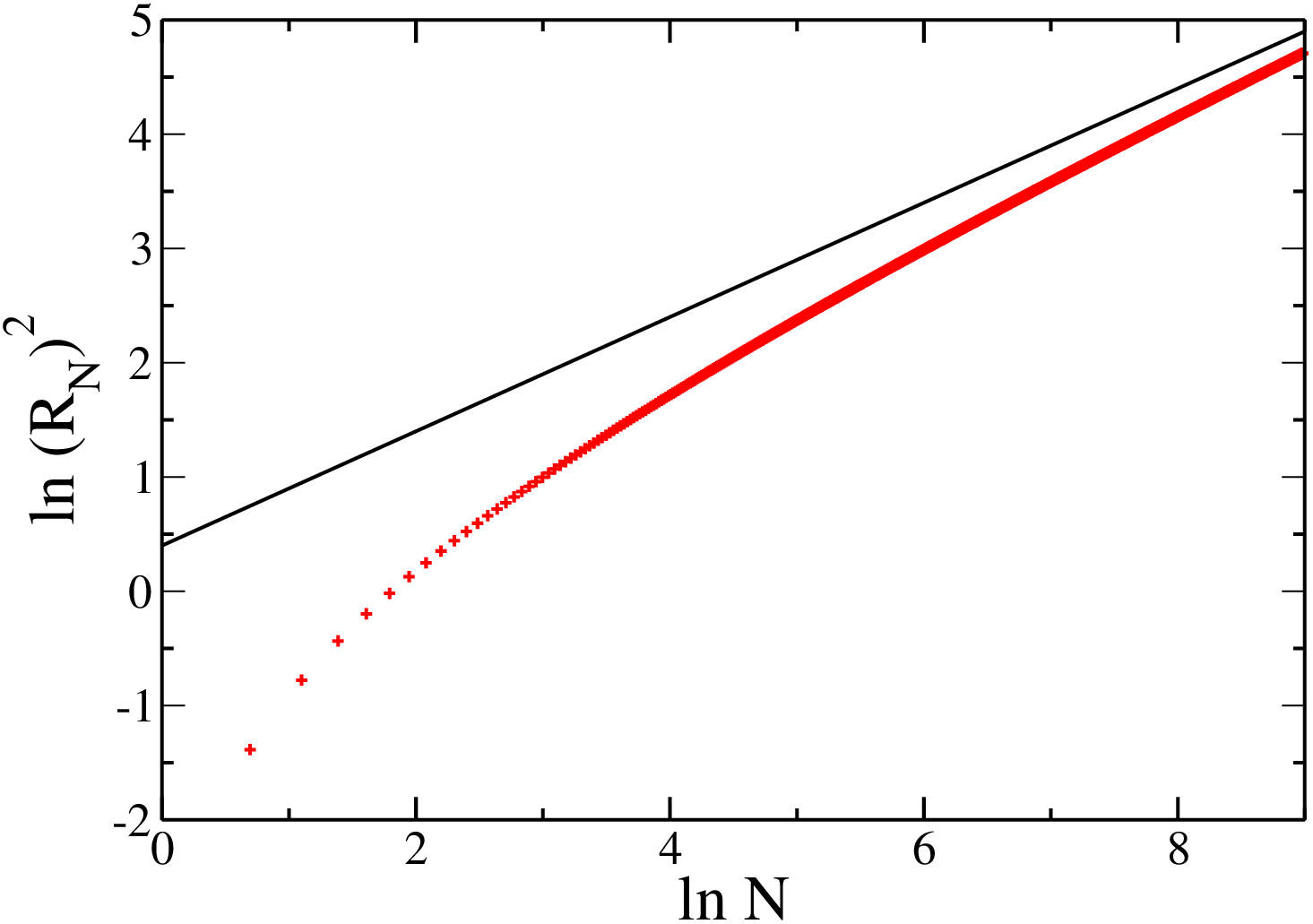}
\caption{The leading dependency of the squared radius of gyration $R_N^2$ on the animal size $N$.
 The line is of slope $2\nu = 1/2$ to guide the eye.
}
\label{Rn}
\end{center}
\end{figure}

\begin{figure}[!t]
\begin{center}
\includegraphics[width=11cm,clip]{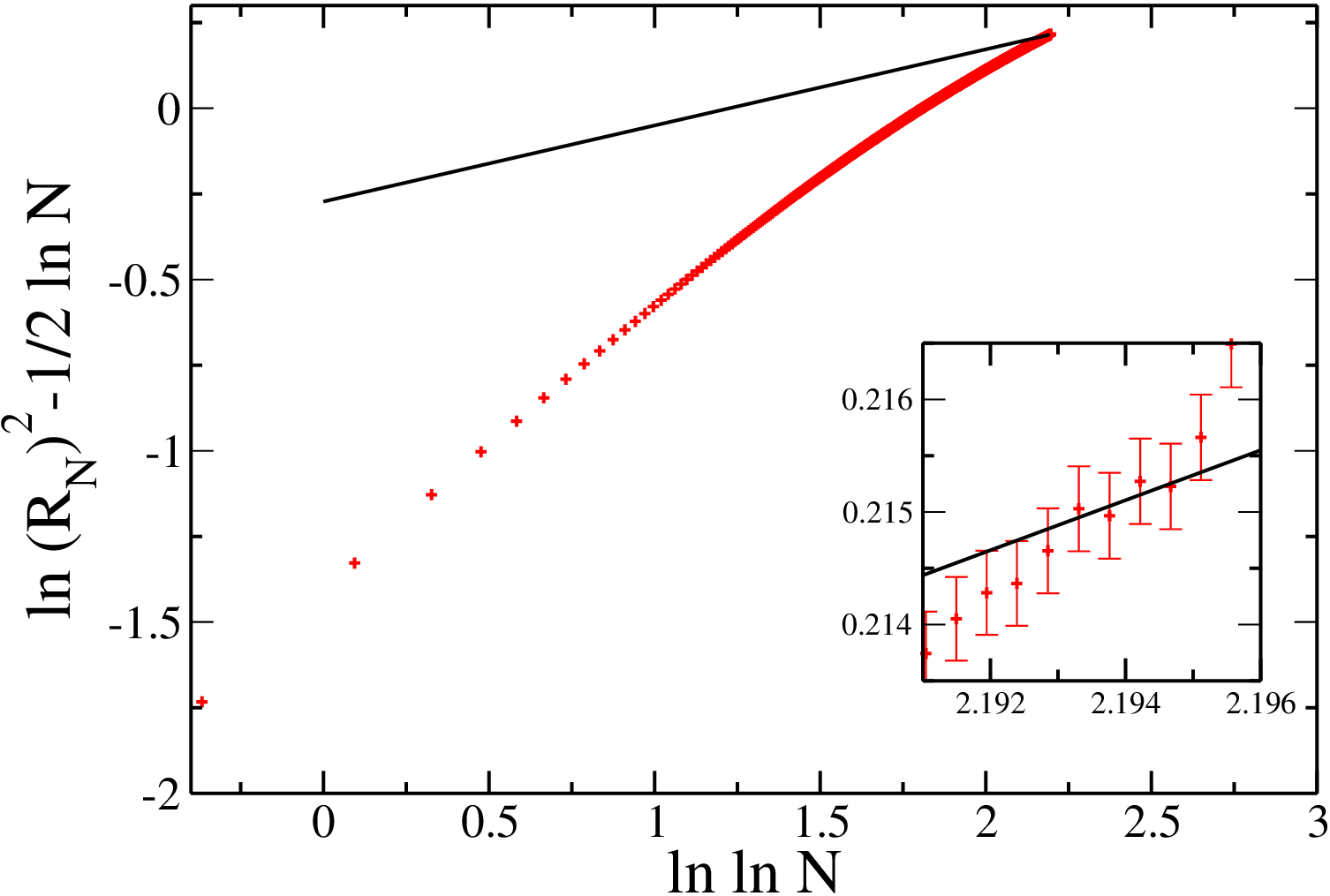}
\caption{The curve identifying the effective logarithmic corrections 
for the gyration radius $R_N^2$, while not asymptotic, is 
of positive slope.
The insert is a zoom for the 10 largest animals and the 
the line is of slope $2\hat{\nu} = 2/9$ to guide the eye.
}
\label{Rnlog}
\end{center}
\end{figure}

The situation for $\hat{\nu}$ is less clear. 
Although the leading behaviour is again well verified, as shown by Fig.~\ref{Rn}, 
the curve for the logarithmic correction is far from having reached its asymptote. The insert in Fig.~\ref{Rn} compares the last portion of the graph with the prediction $\hat{\nu}=1/9$. Whilst tending towards the correct value in the insert, the
true asymptotic value remains to be determined. However, the graph does indicate that the logarithmic correction   
exponent $\hat\nu$ is likely positive, and so supports $\hat\nu=1/9$ more than $\hat\nu=-1/72$.

\section{Discussion}

The scaling relation $\theta=(d-2)\nu+1$,
introduced in Ref.~\cite{PaSo81}, is essentially hyperscaling with the dimension replaced
 by $d-2$ for lattice animals corresponding to dimensional reduction arising in the mapping from lattice animals to the Yang-Lee model. 
In Ref.~\cite{KeJo06}, a set of scaling relations
for logarithmic corrections were developed, which included
the corresponding hyperscaling relation $\hat{\alpha} = d (\hat{q} - \hat{\nu})$
in which $\hat{\alpha}$ is the correction exponent for
the specific heat or free energy and $\hat{q}$ is
a logarithmic-correction exponent for the finite-size scaling of
the correlation length.

In the case of lattice animals, $\hat{\alpha}$ may be identified with $\hat{\theta}$,
from Eq.(\ref{ZZ}). Then, reducing the dimensionality appropriately, we find
\begin{equation}
 \hat{\theta} = (d-2) (\hat{q} - \hat{\nu})
\,.
\label{PSlog}
\end{equation}
This is the logarithmic counterpart to the
Parisi-Sourlas equation. 

The value $\hat{q}=1/6$ was proposed in Ref.~\cite{KeJo06}
for lattice animals, also on the basis of scaling relations.
Together with the 
estimates  $\hat{\theta}=1/3$ and $\hat{\nu}=1/9$,
Eq.(\ref{PSlog}) holds in the present case.
Indeed, the value  $\hat{\nu}=1/9$
fits the full set of scaling relations for logarithmic corrections
proposed in Ref.~\cite{KeJo06}. 
On the other hand, 
 $\hat{\nu}=-1/72$ does not satisfy the scaling relations when used in conjunction with the other exponent values known and reported in Ref.~\cite{RL}.

We have revisited the problem of lattice animals at the upper critical dimension $d=8$
and re-verified that the universal exponents $\theta$ and $\nu$
take their mean-field values there.
We also provide numerical evidence in support of the (uncontested) 
logarithmic counterpart to the
entopic index, $\hat{\theta}=1/3$ and give an estimate for the growth
non-universal constant $\mu$.
Regarding the logarithmic counterpart of the $\nu$ exponent, there are two 
candidate values in the literature. One of these is a direct 
constant-field calculation in Ref.~\cite{RL}, and the other 
is a Fisher renormalized version of the constant-order-parameter 
value also determined in Ref.~\cite{RL}. Only the latter is consistent with the scaling relations,
including a logarithmic version of the Parisi-Sourlas relation proposed herein.
Our numerical evidence also indicates that the latter value is more likely to be the correct one, but, because 
of difficulties in achieving the asymptotic scaling regime, does not 
absolutely confirm the particular value.


\bigskip
%

\end{document}